\documentclass[11pt]{article}

\usepackage{fullpage, hyperref,bbm,amssymb,amsfonts,amsthm}
\usepackage[noend]{algorithmic}
\usepackage{algorithm}
\usepackage{algorithm}
\usepackage{setspace}
\usepackage{color}

\newcommand{\<}{\langle}
\renewcommand{\>}{\rangle}
\newcommand{\C}{\mathbb{C}}
\newcommand{\cA}{\mathcal{A}}

\newcommand{\cH}{\mathcal{H}}

\newtheorem{fact}{Fact}

\newtheorem{definition}{Definition}
\newtheorem{theorem}{Theorem}
\newtheorem{lemma}{Lemma}

\newtheorem{cor}{Corollary}

\begin{document}
\title{Hitting Time of Quantum Walks with Perturbation} 
\author{ Chen-Fu Chiang\thanks{School of Electrical Engineering
and Computer Science,
University of Central Florida, Orlando, FL~32816, USA. Email:
\texttt{cchiang@eecs.ucf.edu}}
\quad
 Guillermo Gomez \thanks{School of Electrical Engineering and
 Computer Science, University of Central Florida, Orlando, FL~32816, USA. Email:
 		\texttt{guillermo.gomez@knights.ucf.edu}}	
}		
		
\maketitle

\begin{abstract}
The hitting time is the required minimum time for a Markov chain-based walk (classical
or quantum) to reach a target state in the state space. 
We investigate the effect of the perturbation on the hitting time of a
quantum walk. We obtain an upper bound for the perturbed quantum
walk hitting time by applying Szegedy's work and the perturbation bounds with
Weyl's perturbation theorem on classical matrix.  Based on the definition of
quantum hitting time given in MNRS algorithm, we further compute the delayed
perturbed hitting time (DPHT) and delayed perturbed quantum hitting time (DPQHT). 
We show that the upper bound for DPQHT is actually greater than the
difference between the square root of the upper bound for a perturbed random
walk and the square root of the lower bound for a random walk.
\end{abstract}

%%%%%%%%%%%%%%%%%%%%%%%%%%%%%%%%%%%%%%%%%%%%%%%%%%%%%%%%%%%%%%%%%%%%%%%%%%%%%%%%%%%%
\section{Introduction}
Markov chains and random walks have been useful tools in classical
computation. One can use random walks to obtain the final stationary 
distribution of a Markov chain to sample from. In such an application the 
time the Markov chain takes to converge, i.e., {\em convergence time}, is
of interest because shorter convergence time means lower cost in
generating a sample. Sampling from stationary distributions of Markov chains combined with simulated 
annealing is the core of many clever classical approximation algorithms. 
For instance, approximating the volume of convex bodies \cite{LV:06}, approximating the 
permanent of a non-negative matrix \cite{JSV:04}, and the partition function of
statistical physics models such as the Ising model \cite{JS:93} and the Potts model \cite{BSVV:08}. 
In addition, one can also use the random walks to search for the {\em{marked}}
state, in which the {\em hitting time} is of interest because hitting time indicates 
the time it requires to find the marked state. \\

In comparison to classical random walk, quantum walk provides a quadratic
speed-up in hitting time. Quantum walk has been applied to solving many
interesting problems \cite{Santha:08}, such as searching problems, group commutativity, 
element distinctness, restricted range associativity, triangle finding in a
graph, and matrix product verification. Perturbations of classical Markov chains are widely studied with respect to
hitting time and stationary distribution. Since a quantum system is susceptible
to the environmental noise, we are interested to know what effect perturbation
has on currently existing quantum walk based algorithms. \\

This work is organized as follows. In section \ref{Deviation} we present the
deviation effect of perturbation on the spectral gap of a classical Markov chain. In
section \ref{HT} we discuss how the hitting time would be affected because of
the perturbation. We explore the upper bounds for the perturbed hitting
time quantumly and the time difference (delayed perturbed hitting time)
both quantumly and classically. Finally in section \ref{conclusion}, we make
our conclusion.

%%%%%%%%%%%%%%%%%%%%%%%%%%%%%%%%%%%%%%%%%%%%%%%%%%%%%%%%%%%%%%%%%%%%%%%%%%%%%%%%%%%%%%%%%
\section{Classical Spectral Gap Perturbation}\label{Deviation}
Given a stochastic symmetric matrix $P \in \C^{n \times n}$, we can quantize the
Markov chain \cite{Szegedy:04}. P.~Wocjan, D. Nagaj and one of us showed that
the implementation of one step of quantum walk \cite{CNW:10} can be achieved efficiently. However,
the above settings always are under the assumption of perfect
scenarios. In real life there are many sources of errors that would perturb
the process. Noise might be propagated along with the input source or
they might be introduced during the process. Here we look solely at the noise
that are introduced at the beginning of the process. \\

The noise can be introduced due to the precision limitation and the noisy
environment. For instance, not all numbers have a perfect binary
representation and the approximated numbers would cause perturbation. Suppose 
our input decoding mechanism can always take the input matrix and represent it
in a symmetric transition matrix $Q$, where $Q$ can be perfectly represented 
and this is the matrix closest to the original matrix $P$
that the system can prepare. \\

Let $E$ be the noise that is introduced because of system's precision
limitation and the environment, we can express the transition matrix as 
\begin{equation}
   Q =  P + E. 
\end{equation} 
\noindent

%***** Please use Gerschgorin Circle Theorem *****\\ 
Classically, much research \cite{IN:09, CM:01, GL:96, Parlett:98,
BF:60,EI:99} has focused on the spectral gaps and stationary distributions 
of the matrices with perturbation. In a recent work by Ipsen and
Nadler \cite{IN:09} , they refined the perturbation bounds for eigenvalues of
Hermitians. Throughout the rest of the paper, $\| \cdot \|$ always denotes the
$l_2$ norm, unless otherwise specified. Based on their result, we summarized the following:

% Not appropriate
% \begin{equation}
% Gap_i(P) := \min_{i \neq j}|\lambda_i(P) - \lambda_j(P)|, \quad 1\leq i \leq n.
% \end{equation}[Give some background and cite some papers and books here,such
% as \cite{Bhatia:97, EI:99}]

\begin{cor}\label{WeylPerturb}
Suppose $P$ and $Q$ $\in \C^{n \times n}$ are Hermitian symmetric transition
matrices with respective eigenvalues
\begin{equation}
0 < \lambda_n(P) \leq \ldots \leq \lambda_1(P) = 1, \quad \quad 
0 < \lambda_n(Q) \leq \ldots \leq \lambda_1(Q) = 1,   
\end{equation}
and $Q = P + E$, then 
\begin{equation}\label{eqn:WeylPerturb}
\max_{1\leq i \leq n} |\lambda_i(P) - \lambda_i(Q)| \leq \|E \|. 
\end{equation}

\noindent
Furthermore, the spectral gap $\delta$ of $P$ and the spectral gap $\Delta$ of
$Q$ have the following relationship
\begin{equation}
\delta - \|E\| \leq \Delta \leq \delta + \|E\|.
\end{equation}
\end{cor}
\proof
Eq.~(\ref{eqn:WeylPerturb}) is a direct result from the Weyl's
Perturbation Theorem. The {\em Weyl's Perturbation Theorem} bounds the worst-case absolute error between the $i$th exact and
the $i$th perturbed eigenvalues of Hermitian matrices in terms of the
$l_2$ norm \cite{GL:96, Parlett:98}. And since
\begin{equation} 
	1 - \lambda_2(P) = \delta, \quad\quad 1 - \lambda_2(Q) = \Delta,\nonumber
\end{equation} \nonumber
by eq.~(\ref{eqn:WeylPerturb}) we have $|\delta -
\Delta| \leq \|E\|$. Therefore, in general we can bound the perturbed
spectral gap $\Delta$ as
\begin{equation}
\delta - \|E\| \leq \Delta \leq \delta + \|E\|. \nonumber 
\end{equation} \nonumber \qed

Generally speaking, the global norm of $E$ might be very large when the
dimensions $n >> 1$ \cite{Johnstone:01}. However, in our case because $E$ is the
difference between two very close stochastic symmetric matrices, its
global norm would never become large.

%%%%%%%%%%%%%%%%%%%%%%%%%%%%%%%%%%%%%%%%%%%%%%%%%%%%%%%%%%%%%%%%%%%%%%%%%%%%%%%%%%%
\section{Hitting Time of Markov Chain Based Walks}\label{HT}
For the purpose of being complete, we need to cite several definitions and
results used in the MNRS algorithm \cite{MNRS:09} in this section. We
recommend interested readers to reference \cite{MNRS:09} for details. \\

Let $P$ be a reversible and ergodic transition matrix with state space $\Omega$
and positive eigenvalues. Suppose $P$ is column-wise stochastic and $|\Omega| =
n$, then let the Markov chain $(X_1, \ldots, X_n)$ under discussion have a
finite state space $\Omega$ and transition matrix $P$.

\begin{definition} For $x \in \Omega$, denote the {\em hitting time} for $x$
	\begin{equation}
    	HT(P, x) = \min\{t\geq 1: X_t = x\}.
    \end{equation}
$HT(P,x)$ is the expected number of transition matrix $P$ invocations to reach
the state $x$ when started in the initial distribution $\pi$. \\ 
\end{definition} 

\begin{definition}
For an $n \times n$ matrix $P$, $P_{-x}$ denotes the $(n-1) \times (n-1)$ matrix
of $P$ where the row and column indexed by $x$ are deleted. For a vector $v$,
$v_{-x}$ is the vector that omits the $x$-coordinate of $v$. Similarly, suppose
$\{M\} = \{x_1, \ldots, x_m\}$, then $P_{-\{M\}}$ denotes the $(n -m)
\times (n -m)$ matrix of P where the rows and columns indexed by  $x_1, x_2,
\ldots, $and $ x_m$ are deleted. \\
\end{definition} 

\begin{definition}
Denote the vector space $\cH = \C^{|\Omega| \times |\Omega|}$. For a state
$|\psi\> \in \cH$, define $\Pi_{\psi} = |\psi\>\<\psi|$ as the orthogonal
projector onto $Span(|\psi\>)$. Let $\cA = Span(|y\>|p_y\> : y \in \Omega)$ be the
vector subspace of $\cH$ where 
\begin{equation}
	|p_y\> = \sum_{z \in \Omega}\sqrt{p_{zy}}|z\>.  
\end{equation}
\noindent 
${\cA}$ is spanned by a set of mutually orthogonal
states $\{|\psi_i\>: i = 1, 2\ldots, |\Omega| \}$, then let
$\Pi_{\cA} = \sum_i \Pi_{\psi_i}$. Similarly, $\cA_{-x}= Span\{|y\>|p_y\>: y
\in \Omega \backslash \{x\}\}$. \\ 
\end{definition}

\begin{definition}The unitary operation $W(P) = (S \cdot (2\Pi_{\cA} - I))^2$ defined on $\cH$ is
the quantum analog of $P$. Similarly, the unitary operation $W(P,x) = (S
\cdot (2 \Pi_{\cA_{-x}} - I))^2$ defined on $\cH$ is the quantum analog of $P_{-x}$. $S$ is the
swap operation defined by $S|y\>|z\> = |z\>|y\>$. \\
\end{definition}

\begin{fact}\label{fact:WalkToUnitary}\cite{MNRS:09, Szegedy:04}
Let $x \in \Omega$ and $|\mu\> = |x\>|p_x\>$. Let $U_2 =
S(2\Pi_{\cA}-I)$ and $U_1 = I - 2|\mu\>\<\mu|$. When $P$ is reversible, then
$U_2^2 = W(P)$ and $(U_2U_1)^2 = (S(2\Pi_{\cA_{-x}} -I))^2 = W(P,x)$.
\proof
Since $\Pi_{\cA} = \Big(\sum_{y=1, y\neq x}^{|\Omega|}|y\>|p_y\>\<y|\<p_y|\Big)
+ |\mu\>\<\mu|$, then we have 
\begin{eqnarray*}
U_2U_1 & = & S(2\Pi_{\cA} -I)(I - 2|\mu\>\<\mu|) \\
	   & = & S(2 \Pi_{\cA} -2|\mu\>\<\mu| - I) \\
	   & = & S(2 \Pi_{\cA_{-x}} - I) 
\end{eqnarray*}
\end{fact}

%%%%%%%%%%%%%%%%%%%%%%%%%%%%%%%%%%%%%%%%%%%%%%%%%%%%%%%%%%%%%%%%%%%%%%%%%%%%
\subsection{Classical Hitting Time}\label{ClassicalHittingTime}
By \cite{MNRS:09}, the $x$-hitting time of $P$ can be expressed as $HT(P,x) =
\pi^\dagger (I-P_{-x})^{-1}u_{-x}$, where $u$ is an all-ones vector. It is known that
\begin{equation}
	\pi_{-x}^\dagger (I - P_{-x})^{-1} u_z = \sqrt{\pi_{-x}}^\dagger(I -
	S_{-x})^{-1}\sqrt{\pi_{-x}}
\end{equation} 
where $S_{-x} = \sqrt{\Pi_{-x}} P_{-x} \sqrt{\Pi_{-x}}^{-1}$ with 
$\Pi_{-x} = diag({\pi_i})_{i \neq x}$ and $\sqrt{\pi_{-x}} $ is  the
entry-wise square root of $\pi_{-x}$. Let $\{v_j: j \leq n -1 \}$ be the
set of normalized eigenvectors of $S_{-x}$ where the eigenvalue of $v_j$ is $\lambda_j = \cos \theta_j$ 
with $0 \leq \theta_j < \pi/2$. By reordering the
eigenvalues, let us assume that $1 > \lambda_1 \geq \ldots
\geq \lambda_{n-1} > 0$. When $\sqrt{\pi_{-x}} = \sum_j \nu_j v_j$ is the decomposition 
of $\sqrt{\pi_{-x}}$ in the eigenbasis of
$S_{-x}$, the $x$-hitting time satisfies:
\begin{equation}
	HT(P,x) = \sum_j \frac{\nu_j^2}{1 - \lambda_j}.
\end{equation}

\noindent
Two simple facts can be observed from above description of classical hitting time.  
\begin{fact}
$S_{-x}$ and $P_{-x}$ are {\em similar}, they have the {\textbf{same
eigenvalues}}.
\end{fact}

\begin{fact}\label{fact:amplitudesum}
Since the entries of distribution $\pi$ sum up to 1 , i.e. $\sum_i (\pi_i)= 1$, then it is obvious that $\sqrt{\pi_{-x}}^\dagger \sqrt{\pi_{-x}}
= \sum_i (\pi_i)_{i \neq x} \leq 1$. Hence we know that {$\mathbf{\sum_i
\nu_i^2 = \sum_i \tilde{\nu}_i^2 \leq 1}$}. \\
\end{fact}

%%%%%%%%%%%%%%%%%%%%%%%%%%%%%%%%%%%%%%%%%%%%%%%%%%%%%%%%%%%%%%%%%%%%%%%%%%%%%%%% 
\subsection {Delayed Perturbed Hitting Time}\label{SEC:DPHT}
In this subsection, we define the delayed perturbed hitting time and its upper bound as the following.   

	\begin{lemma}
        For a Markov transition matrix $P$ with state
        space $\Omega$ and limiting distribution $\pi$. Assume  
        $|\Omega|=n$ and let $|v_i\>$ be the eigenvector with corresponding eigenvalue
        $\lambda_i$ of $P_{-x}$. Suppose the eigenvalues of $P_{-x}$ are ordered such that $1 > \lambda_1 \geq
        \lambda_2 \geq \ldots \geq \lambda_{n-1} > 0$. The $x$-hitting time satisfies
        \begin{equation}
        	HT(P,x) = \displaystyle \sum_{j=1} \frac{\nu_j^2}{1 -
        	\lambda_j} \nonumber
        \end{equation} 
		where $\sqrt{\pi_{-x}} = \sum_{j=1}^{|\Omega|-1} \nu_j|v_i\>$. When given a
		perturbed matrix $Q$ where $\|Q - P\| \leq \|E\|$  then the
		Delayed Perturbed Hitting Time (DPHT(P,Q,x)) is bounded from above by 
		\begin{equation}
        	 \frac{1}{1-\lambda_1 - \|E\|_2} - \frac{1}{1 - \lambda_{1} + \gamma}
		\end{equation}
		where $\lambda_1 - \lambda_{n-1} = \gamma$.
		
		\proof 
		Let the eigenvalues of $Q_{-x}$ be $\tilde{\lambda}_i$. By the fact $\|Q-P\|
		\leq \|E\|$ and Weyl's perturbation theorem, we know that $\|Q_{-x} - P_{-x}\| \leq \|E\|$ and $|\lambda_i - \tilde{\lambda}_i| \leq
		\|E\|$. The delayed hitting time due to perturbation is thus  
 		\begin{eqnarray}\label{eqn:DPHT}
 			DPHT(P,Q,x) & = & HT(Q,x) - HT(P,x) \nonumber\\
 			&=& \displaystyle \sum_{i \in
 			\Omega}\Big(\frac{\tilde{\nu}_i^2}{1-\tilde{\lambda}_i}	-\frac{\nu_i^2}{1-\lambda_i}\Big) \nonumber\\
 		    %&\leq& \max_i \Big(\displaystyle \sum_{i \in \Omega}\frac{\tilde{\nu}_i^2}{1-\tilde{\lambda}_i}\Big) - \min_i\Big(\displaystyle \sum_{i \in \Omega}\frac{\nu_i^2}{1-\lambda_i}\Big) \\
 			& \leq & \Big(\displaystyle \sum_{i \in \Omega}\frac{\tilde{\nu}_i^2}{1-\lambda_1-\|E\|}\Big) - \Big(\displaystyle \sum_{i \in \Omega}\frac{\nu_i^2}{1-\lambda_{n-1}}\Big)  \nonumber\\
 			&\leq& \big( \frac{1}{1-\lambda_1 - \|E\|} - \frac{1}{1 - \lambda_{1} +	\gamma}\big), 
 			\end{eqnarray}	\qed
        \end{lemma}
		\noindent
		the last inequality is a result from {\em Fact \ref{fact:amplitudesum}}. 

%%%%%%%%%%%%%%%%%%%%%%%%%%%%%%%%%%%%%%%%%%%%%%%%%%%%%%%%%%%%%%%%%%%%%%%%%%%%%%%%%%%%%%%%%
\subsection{Upper Bound for Perturbed Quantum Hitting Time}\label{hittingtimeDeviation}
Given two Hermitian stochastic matrices, $P$ and $Q$, we 
explore the difference between walk operators, $W(P)$ and $W(Q)$, with respect
to their hitting time. Denote the set of marked elements as
$|M|$. Based on the result from {\em Corollary \ref{WeylPerturb}}, we have the
following:

\begin{cor}
Given two symmetric reversible ergodic transition matrices $P$ and
$Q$ $\in \C^{n \times n}$, where $Q = P + E$, let $W(P)$ and $W(Q)$ be
quantum walks based on $P$ and $Q$, respectively. Let $M$ be the set of marked
elements in the state space. Denote $QHT(P)$ as the hitting time of walk $W(P)$
and $QHT(Q)$ as the hitting time of walk $W(Q)$. Suppose $|M| = \epsilon N$. If the
second largest eigenvalues of $P$ and $Q$ are at most $1 - \delta$ and $1-\Delta$, respectively, 
then in general

\begin{equation}
QHT(P) = O\Big(\sqrt{\frac{1}{\delta \epsilon}}\Big), \quad \quad QHT(Q) =
O\Big(\sqrt{\frac{1}{(\delta-\|E\|)  \epsilon}}\Big)
\end{equation}
where $ \delta -\|E\| \leq \Delta \leq \delta + \|E\|$.

\proof Suppose the Markov chain $P$, $Q$ and matrix $E$ are in the following
block structure
\begin{equation}
P = \left( \begin{array}{cc}
P_1 & P_2 \\
P_3 & P_4 \end{array} \right) , \quad
Q = \left( \begin{array}{cc}
Q_1 & Q_2 \\
Q_3 & Q_4 \end{array} \right), \quad
E = \left( \begin{array}{cc}
E_1 & E_2 \\
E_3 & E_4 \end{array} \right)
\end{equation}
where we order the elements such that the marked ones come last, i.e., $P_4$,
$Q_4$ and $E_4 \in \C_{|M| \times |M|}$. The corresponding modified Markov
chains \cite{Szegedy:04} would be
\begin{equation}
\tilde{Q} = \left( \begin{array}{cc}
Q_1 & 0 \\
Q_3 & I \end{array} \right) = \left( \begin{array}{cc}
P_1 + E_1 & 0 \\
P_3 + E_3 & I \end{array} \right).
\end{equation} 

By \cite{Szegedy:04}, we have $QHT(P) = O(\sqrt{\frac{1}{1-\|P_1\|}})$
and $QHT(Q) = O(\sqrt{\frac{1}{1-\|Q_1\|}})$. Since we know
\begin{equation} \|P_1\| \leq 1 - \frac{\delta \epsilon}{2} 
\quad \mbox{and} \quad \|Q_1\| \leq 1 - \frac{\Delta \epsilon}{2} 
\end{equation}
by \cite{Szegedy:04} and by
Cauchy's interlacing theorem we have  $\|E\| \geq \|E_1\|$
\cite[Cor.III.1.5]{Bhatia:97}, we then obtain 
\begin{equation}\|Q_1\| \leq \min\left\{\|P_1\| + \|E\|, 1-\frac{(\delta
-\|E\|)\epsilon}{2}\right\} 
\end{equation}
as $ \delta -\|E\| \leq \Delta \leq \delta + \|E\|$.
Therefore, the hitting times for $P$ and $Q$ are derived.
\qed
\end{cor}

From the corollary above, it is clear that the noise increases the the
quantum hitting time. By a simple comparison with the classical hitting time,
we have the following fact.

\begin{fact}
When given a perturbed quantum walk $W(Q)$, where the magnitude of noise is
$\|E\|$, the quadratic speed-up gained from the quantum walk will be
annihilated when $\|E\| \geq \Omega(\delta(1-\delta \epsilon))$. 
\end{fact}

%%%%%%%%%%%%%%%%%%%%%%%%%%%%%%%%%%%%%%%%%%%%%%%%%%%%%%%%%%%%%%%%%%%%%%%%%%%%%%%%%%%
\subsection{Quantum Hitting Time Based on MNRS Algorithm}\label{SEC:QHTMNRS}
Let $U=U_2U_1$ be a unitary matrix with real entries. Let $|\mu\>$ (see Fact
\ref{fact:WalkToUnitary}) be the marked element where $U_1=I - 2|\mu\>\<\mu|$ and $U_2$ is a real unitary matrix with a
unique 1-eigenvalue $|\phi\>$. Similar to the classical case, let
$|\phi\>_{-\mu} = |\phi\> - \<\phi|\mu\>|\mu\>$. \\

The potential eigenvalues for $U$ are then $\pm 1$ and conjugate complex numbers
$(e^{i\alpha_j}, e^{-i\alpha_j})$. Let $|\phi\>_{-\mu}$ be the input state for the phase
estimation of $U$, then $|\phi\>_{-\mu}$ can be uniquely decomposed in the eigenbasis of $U$ as
\begin{equation}
	|\phi\>_{-\mu} =  \delta_0|\omega_0\> + \sum_{j} \delta_j |\omega_j^{\pm}\>  +
	\delta_{-1}|\omega_{-1}\>
\end{equation}
where $U|\omega_0\> = |\omega_0\>$, $U|\omega_{-1}\> = -|\omega_{-1}\>$ and
$U|\omega_j\> = e^{\pm i\alpha_j}|\omega_j\>$. Let $QH$ be the
random variable which takes the value $1/\alpha_j$ with probability $\delta_j^2$
and the value $1/\pi$ with probability $\delta_{-1}^2$. 

\begin{definition}\label{def:QHitTime}\cite{MNRS:09} The quantum $|\mu\>$-hitting time of $U_2$ is the expectation of QH, that is
\begin{equation}
QHT(U_2, |\mu\>) = 2 \sum_{i}\frac{\delta_j^2}{\alpha_j}+ \frac{\delta_{-1}^2}{\pi}.  
\end{equation}
\end{definition}
Hence, in order to compute the quantum hitting time of $U_2$, it is important
to compute the spectral decomposition of $U$. It is shown in the following
theorem.

\begin{theorem}\label{thm:Qspectrum}\cite{Szegedy:04}
Fix an $n \times n$ column-wise stochastic matrix $\tilde{P}$, and let
$\{|\lambda\>\}$ denote a complete set of orthonormal eigenvectors of the $n \times n $ matrix $D$ with
entries $D_{jk} = \sqrt{\tilde{P}_{jk} \tilde{P}_{kj}}$ with eigenvalue
$\{\lambda\}$. Then the eigenvalues of the discrete-time quantum walk $U = S(2\Pi_{\cA} - I)$
corresponding to $\tilde{P}$ are $\pm 1$ and $\lambda \pm i\sqrt{1 - \lambda^2}
= e^{\pm i arccos \lambda}$ \footnote{Eigenvalues of $\tilde{D}$ are exactly the eigenvalues of
$\tilde{P}_{-\{M\}}$ and eigenvalue 1.}.
\end{theorem}

Let the subset $M$ be the set of marked elements that we are searching for. 
The discrete-time quantum walk $U_{-\{M\}} = S(2 \Pi_{\cA_{-\{M\}}} -I)$
satisfies the above theorem when we modify the original transition matrix $P$
into $\tilde{P}$ in the following manner:
 		\[	
 		 \tilde{P}_{jk}  = \left\{ 
  			 \begin{array}{l l l}
 	   		  1 & \quad & \textrm {$k \in M$ and $j = k$}\\
     		  0 & \quad & \textrm {$k \in M$ and $j \neq k$}\\
     		  P_{jk} & \quad & \textrm {$k \not\in M$}\\
 		  \end{array} \right.	
 		\]		
		\noindent 
		We can view $\tilde{P}$ in block structure as follows:		
		\begin{equation}
        P = \left( \begin{array}{cc}
		P_{-\{M\}} & P_2 \\
		P_3 & P_4 \end{array} \right) \longrightarrow \quad
        \tilde{P} = \left( \begin{array}{cc}
					P_{-\{M\}} & 0 \\
					P_3 & I \end{array} \right), 
        \end{equation}
		
		\noindent 
	    then the corresponding discriminant matrix $\tilde{D}$
		is \begin{equation}
        \tilde{D} = \left( \begin{array}{cc}
					P_{-\{M\}} & 0 \\
					0 & I \end{array} \right) . 
        \end{equation}
% 		
% 		\begin{lemma}\label{lem:LGPM} \cite{Szegedy:04}
%         If the second largest eigenvalue of $P$ (in absolute value) is at most
%         $1 - \delta$ and $|M| \leq \epsilon N$, then $\|P_{-\{M\}}\| \leq 1 -
%         \frac{\delta \epsilon}{2}$. \\ 
%         \end{lemma}
% 	   
% \begin{cor}
% Let $\tilde{\delta}$ be the spectral gap of $\tilde{D}$ and $\epsilon
% =|M|/N$ where $M$ is the set of marked elements we wish to detect (or find). 
% Let $\delta$ be the spectral gap of the original matrix $P$ and then we know
% that $\tilde{\delta} \geq \frac{\delta \epsilon}{2}$.
%  
% \proof $\tilde{\delta}= 1 - \|P_{-\{M\}}\| \geq \frac{\delta \epsilon}{2} $.\qed
% \end{cor}

\begin{fact}
Now let us set $M = \{x\}$. Then $\pm 1$ and $e^{\pm i \alpha_j}$ are
eigenvalues of $U_{-x}$ where $\lambda_j$ are the eigenvalues of $P_{-x}$. Since
$\lambda_j = \cos \theta_j$ (see sect. \ref{ClassicalHittingTime}), and by use of {\em theorem \ref{thm:Qspectrum}}, 
we know that $\theta_j = \alpha_j$. 
\end{fact}	
	
	Furthermore, by {\em Fact \ref{fact:WalkToUnitary}} we know the unitary
	$W(P,x) = U_{-x}^2$. The eigenvectors of $U_{-x}$ remain the eigenvectors of
	$W(P,x)$ but the eigenvalues of $W(P,x)$ would be $e^{2i\alpha_j}$. Given $|\phi\>_{-\mu}$ as the input state, 
	we run phase estimation of $W(P,x)$ and the corresponding
	quantum hitting time would be 
	\begin{equation}\label{eqn:QHitTime}
	QHT(P,x) = 2\sum_{j=1}^{n-1}\frac{\delta_j^2}{2\alpha_j} =  \sum_{j=1}^{n-1}\frac{\delta_j^2}{\theta_j},  
	\end{equation} 
	the term $\frac{\delta_{-1}^2}{\pi}$ in {\em def. \ref{def:QHitTime}} disappears because the corresponding eigenphase becomes 0. 

%%%%%%%%%%%%%%%%%%%%%%%%%%%%%%%%%%%%%%%%%%%%%%%%%%%%%%%%%%%%%%%%%%%%%%%%%%%%%%%%%%%
	\subsection{Delayed Perturbed Quantum Hitting Time}\label{SEC:DPQHT}	
	In this subsection, we define the Delayed Perturbed Quantum Hitting Time (DPQHT) and its upper bound as the following.  
	\begin{fact}\cite{MNRS:09}\label{QHTFact}
	When $P$ is an ergodic Markov transition with positive eigenvalues, then the $x$-quantum hitting time for the unitary $W(P,x)$ is
	\begin{equation}\label{eqn:QHT}
	QHT(P, x) = \sum_{j=1}^{n-1}\frac{\nu_j^2}{\theta_j}
	\end{equation}
	\proof Since the length of the projection of $|\phi\>_{-u}$ to the eigenspace corresponding to $\alpha_j$ is $\nu_j^2$ \cite{MNRS:09},  
	then by eq.~\ref{eqn:QHitTime} we have the result as shown in eq.~\ref{eqn:QHT}.  
	\end{fact}

	\begin{lemma} 
    Given $QHT(P,x)$ and $QHT(Q,x)$ with $\|P - Q\| = \|E\|$, then by use of
    Fact \ref{QHTFact}, we have the Delayed Perturbed Quantum Hitting Time
    $DPQHT(P,Q,x)$ bounded from above by 
    \begin{eqnarray*}
    \frac{1}{\sqrt{1-\lambda_1 - \|E\|_2}} - \frac{1}{2
    \sqrt{1-\lambda_1 + \gamma}}. 
    \end{eqnarray*}	
	The eigenvalues of $P_{-x}$ are ordered such that 
	$1 > \lambda_1 \geq  \lambda_2 \geq \ldots \geq \lambda_{n-1} > 0$
	and $\lambda_1 - \lambda_{n-1} = \gamma$.
    \proof
    Based on Fact \ref{QHTFact}, we have
    \begin{eqnarray}\label{eqn:DPQHT}
    DPQHT(P,Q,x) & = & QHT(Q,x) - QHT(P, x)  \nonumber\\ 
    & = & \sum_{i\in \Omega} \Big(\frac{\tilde{\nu}_i^2}{\tilde{\theta}_i} -\frac{\nu_i^2}{\theta_i}\Big) \nonumber\\    
    %& \leq & \max_i\Big(\sum_{i\in \Omega}\frac{\tilde{\nu}_i^2}{\tilde{\theta_i}}\Big) - \min_i \Big(\sum_{i\in \Omega}\frac{\nu_i^2}{\theta_i}\Big) \\
    & \leq & \Big(\sum_{i\in \Omega}\frac{\tilde{\nu}_i^2}{\cos^{-1}\tilde{\lambda}_1}\Big) - \Big(\sum_{i\in \Omega}\frac{\nu_i^2}{\cos^{-1}\lambda_{n-1}}\Big) \nonumber\\
    & \leq & \frac{1}{\sqrt{1-\lambda_1 - \|E\|}} - \frac{1}{2 \sqrt{1-\lambda_1 + \gamma}}. 
    \end{eqnarray}  \qed
     \end{lemma}
	\noindent
	The last inequality is a simple result from {\em Fact \ref{fact:amplitudesum}}
	and the fact that $2\sqrt{1-\lambda} > \cos^{-1}\lambda > \sqrt{1-\lambda}$ for
	all $\lambda \in (0,1)$. \\

\section{Conclusion}\label{conclusion}
By quantizing a perturbed symmetric stochastic $n \times n$ matrix $Q$ with noise $E$, we
find an upper bound for the perturbed quantum hitting time. We also so show the lower bound for the magnitude of 
noise when the quadratic speed-up gained from the quantum walk will be
annihilated by the noise. \\

Furthermore we compute the upper bound for the delayed perturbed quantum hitting time based 
on the definition of quantum hitting time. One cannot just directly apply
the square root speed-up from quantum walk to the delayed perturbed hitting
time (see eq.~\ref{eqn:DPHT}). If one does so, one would obtain an upper bound
for $DPQHT$ as
\begin{equation}\label{eqn:wrongDPQHT}
 \frac{1}{\sqrt{1-\lambda_1 - \|E\|}} - \frac{1}{\sqrt{1-\lambda_1 + \gamma}}.
\end{equation}

\noindent
It would be incorrect. The second term of eq.~\ref{eqn:wrongDPQHT} should the
the minimum of $ \sum_{i\in \Omega}\frac{\nu_i^2}{\cos^{-1}\lambda_{n-1}}$. But in
eq.~\ref{eqn:wrongDPQHT}, the second term was actually the maximum. 
Thus, it is clear that the upper bound for DPQHT is actually greater than the difference between the square root of the upper bound 
for a perturbed random walk and the square root of the lower bound for a random walk.
%%%%%%%%%%%%%%%%%%%%%%%%%%%%%%%%%%%%%%%%%%%%%%%%%%%%%%%%%%%%%%%%%%%%%
\section{Acknowledgments}
		C.~C. gratefully acknowledges the support of NSF grants
		CCF-0726771 and CCF-0746600. G.~G gratefully acknowledges the support 
		of NSF REU supplement to grant CCF-0726771. We would also like to thank 
		H. Ahmadi for his useful comments.
%%%%%%%%%%%%%%%%%%%%%%%%%%%%%%%%%%%%%%%%%%%%%%%%%%%%%%%%%%%%%%%%%%

%\include{Bib}
%%%%%%%%%%%%%%%%%%%%%%%%%%%%%%%%%%%%%%%%%%%%%%

\begin{thebibliography}{10}
% 	 \bibitem{KMOR:10}
%      	H.~Krovi, F.~Magniez, M.~Ozols, and J.~Roland, 
%      	{\em Finding is as Easy as Detecting for Quantum Walks}. In 37th
%      	International Colloquium on Automata, Languages and Programming
%      	(ICALP'10), vol.~6198, Lecture Notes in Computer Science, pp. 540-551, 
%      	Springer, 2010
	 \bibitem{LV:06}
		L.~Lov\'{a}sz and S.~Vempala, {\em Simulated Annealing in Convex Bodies and an $O^*(n^4)$ Volume Algorithm},
		Journal of Computer and System Sciences, vol.~72, issue 2, pp. 392--417, 2006.

	\bibitem{JSV:04}
		M.~Jerrum, A.~Sinclair, and E.~Vigoda, {\em A Polynomial-Time
		Approximation Algorithm for the Permanent of a Matrix Non-Negative
		Entries}, Journal of the ACM, vol.~51, issue 4, pp.~671--697, 2004.

	\bibitem{JS:93}
		M.~Jerrum and A.~Sinclair, {\em Polynomial-Time Approximation
		Algorithms for the Ising Model}, SIAM Journal on Computing, vol.~22,
		pp.~1087--1116, 1993.

	\bibitem{BSVV:08}
		I.~Bez\'{a}kov\'{a}, D.~\v{S}tefankovi\v{c}, V.~Vazirani and E.~Vigoda,
		{\em Accelerating Simulated Annealing for the Permanent and
		Combinatorial Counting Problems}, SIAM Journal on Computing, vol.~37,
		No. 5, pp.~1429--1454, 2008.
     	     	
	\bibitem{Santha:08}
		M.~Santha, {\em Quantum Walk Based Search Algorithms}, Proc. of 5th Theory and
		Applications of Models of Computation (TAMC08), Lectures Notes on Computer
		Science, vol. 4978, pp. 31 - 46, 2008 
		
		 \bibitem{Szegedy:04}
		M.~Szegedy, {\em Quantum Speed-up of Markov Chain Based Algorithms},
		Proc. of 45th Annual IEEE Symposium on Foundations of Computer
		Science, pp.~32--41, 2004.
		
	\bibitem{CNW:10}
		C.~Chiang, D.~Nagaj, P.~Wocjan, {\em Efficient Circuits for the Quantum
		Walks}, QIC vol.~10 no.~5\&6 pp.~0420--0434, 2010.
			
	\bibitem{IN:09}
        I.~Ipsen and B.~Nadler, {\em Refined Perturbation Bounds for Eigenvalues
        of Hermitian and Non-Hermitian Matrices}, SIAM J. Matrix Anal. Appl.,
        vol.~31, no.~1, pp.~40--53, 2009.

    \bibitem{CM:01}
       G.~Cho and C.~Meyer, {\em Comparison of Perturbation Bounds for the
       Stationary Distribution of a Markov Chain}, vol.~335, issue 1-3, pp.137 -
       150, Linear Algebra and Its Applications, 2001.
		
	\bibitem{GL:96}
       G.~Golub and C.~Loan, {\em Matrix Computations}, 3rd ed., The Johns
       Hopkins University Press, 1996. 
       
    \bibitem{Parlett:98}
       B.~Parlett, {\em The Symmetric Eigenvalue Problems}, SIAM, Philadelphia,
       1998. 
       
    \bibitem{BF:60}
        F.~Bauer and C.~Fike, {\em Norms and Exclusion Theorems}, Numer.
        Math.,vol.~2, pp.~137 - 141, 1960. 	
        
      \bibitem{EI:99}
        S.~Eisenstat and I.~Ipsen, {\em Three Absolute Perturbation Bounds for
        Matrix Eigenvalues Imply Relative Bounds}, SIAM Journal on Matrix
        Analysis and Applications, vol.~20 ,  issue 1, pp.~149 - 158, 1999.
		
	 \bibitem{Johnstone:01}
       I.~Johnstone, {\em On the Distribution of the Largest Eigenvalue in
       Principal Components Analysis}, vol.~29, no.~2, pp.~295 - 327, Annals of
       Statistics, 2001.
		
	\bibitem{MNRS:09}
     	F.~Magniez, A.~Nayak, P.~Richter and M.~Santha, {\em On the Hitting Times
     	of Quantum versus Random Walks}, Proc. of the twentieth
     	annual ACM-SIAM Symposium on Discrete Algorithms, pp. 86 - 95, 2009.
%      	
%      \bibitem{Chiang:10}
%      	C.~Chiang, {\em Sensitivity of Quantum Walks with Perturbation}, 
%      	Proc. of the 10th Asian Conference on Quantum Information Science, pp.
%      	209-210, 2010.
     	

	
% 	\bibitem{WA:08}
% 		P.~Wocjan and A.~Abeyesinghe, {\em Speed-up via Quantum Sampling},
% 		Phys. Rev. A, vol.~78, pp.~042336, 2008.
% 	

    
    \bibitem{Bhatia:97}
        R. Bhatia, {\em Matrix Analysis}, Springer Verlag, New York, 1997.     
        
%     \bibitem{SBB:07}
% 		R.~Somma, S.~Boixo, and H.~Barnum, {\em Quantum Simulated Annealing},
% 		arXiv:abs/0712.2008
% 		
% 	\bibitem{SBBK:08}
% 		R.~Somma, S.~Boixo, H.~Barnum, E.~Knill, {\em Quantum Simulations of
% 		Classical Annealing Processes}, Phys. Rev. Lett. vol.~101, pp.~130504, 2008.
%        

   
\end{thebibliography}
	 \end{document}